\documentclass[tightenlines,showpacs,prx,floatfix,preprintnumbers,amsmath,superscriptaddress,
amssymb,amsbsy,nofootinbib,nobibnotes,twocolumn,longbibliography,prx]{revtex4-1}

\usepackage{amsmath, amssymb, amsfonts,amsbsy,amsthm} 
\usepackage{graphicx,color} 
\usepackage{acronym}
\usepackage{units,subfigure} 
\usepackage{hyperref} 
\usepackage{bm}
\usepackage{times}
\usepackage{mathrsfs}
\usepackage{latexsym}
\usepackage{dsfont}
\usepackage{hyphenat} 
\usepackage{titlesec}
\usepackage{enumitem} 

\definecolor{darkgreen}{rgb}{0.0, 0.5, 0.0}
\definecolor{purple}{rgb}{0.56, 0.0, 1.0}

\newcommand{\gw}{_\textsc{gw}}
\newcommand{\bbh}{_\textsc{bbh}}
\newcommand{\ml}{^\textsc{ml}}

\newcommand{\plaw}{_\textsc{pl}}
\newcommand\intHalf{\int_0^\infty} 
\newcommand{\Msol}{M_\odot}

\acrodef{BNS}{Binary Neutron Star}
\acrodef{BBH}{Binary Black Hole}
\acrodef{SNR}{signal-to-noise ratio}

\begin{document}
\title{Limits of Astrophysics with Gravitational Wave Backgrounds}

\author{Thomas Callister}
\email{tcallist@caltech.edu}
\affiliation{LIGO Laboratory, California Institute of Technology, MS 100-36,
Pasadena, California, 91125, USA}

\author{Letizia Sammut}
\affiliation{School of Physics and Astronomy, Monash University, Clayton, Victoria 3800, Australia}

\author{Shi Qiu}
\affiliation{School of Physics and Astronomy, Monash University, Clayton, Victoria 3800, Australia}

\author{Ilya Mandel}
\affiliation{School of Physics and Astronomy, University of Birmingham, Birmingham B15 2TT, United Kingdom}

\author{Eric Thrane}
\affiliation{School of Physics and Astronomy, Monash University, Clayton, Victoria 3800, Australia}

\date{\today}

\begin{abstract}

The recent Advanced LIGO detection of gravitational waves from the binary black hole GW150914 suggests there exists a large population of merging binary black holes in the Universe. 
Although most are too distant to be individually resolved by advanced detectors, the superposition of gravitational waves from many unresolvable binaries is expected to create an astrophysical stochastic background.
Recent results from the LIGO and Virgo collaborations show that this astrophysical background is within reach of Advanced LIGO.
In principle, the binary black hole background encodes interesting astrophysical properties, such as the mass distribution and redshift distribution of distant binaries.
However, we show that this information will be difficult to extract with the current configuration of advanced detectors (and using current data analysis tools).
Additionally, the binary black hole background also constitutes a foreground that limits the ability of advanced detectors to observe other interesting stochastic background signals, for example from cosmic strings or phase transitions in the early Universe. 
We quantify this effect.

\end{abstract}

\maketitle

\section{INTRODUCTION} \label{intro}

The first direct detection of gravitational waves was recently announced by the Advanced LIGO (Laser Interferometer Gravitational Wave Observatory) and Virgo Collaborations~\cite{gw150914,cqg.32.074001.15,gw150914_aLIGO,cqg.32.024001.15}.
The observation of GW150914, a binary black hole (BBH) merger with individual black hole masses of $36$ and 29$\Msol$ at a luminosity distance of $\approx \unit[400]{Mpc}$~\cite{gw150914_pe}, implies that the masses and coalescence rate of stellar-mass BBHs are at the high end of previous predictions~\cite{gw150914_rate,gw150914_astro,CBC_2010_rates}.
As a consequence, the astrophysical stochastic gravitational wave background, arising from all coalescing binary black holes too distant to individually resolve~\cite{2011ApJ...739...86Z,2011PhRvD..84h4004R,2011PhRvD..84l4037M,2013PhRvD..87d2002W,2013MNRAS.431..882Z,2015A&A...574A..58K}, is potentially within reach of advanced detectors.
When operating at design sensitivity, Advanced LIGO may detect this binary black hole background with signal-to-noise ratio $\mathrm{SNR}=3$ in as few as 1.5 years~\cite{gw150914_stoch}.
However, there is significant uncertainty in the strength of the stochastic signal due to uncertainty in the coalescence rate, currently estimated from only 16 days of double-coincident observation ~\cite{gw150914_stoch,gw150914_rate}.
In this paper, we build on the LIGO and Virgo results from Ref. \cite{gw150914_stoch} and investigate the potential to extract astrophysical information from measurements of the stochastic background.

The detection of an astrophysical stochastic background would be a major accomplishment, providing us with a glimpse of sources at cosmological distances.
Given this exciting possibility, we address three key questions concerning the future prospects for gravitational wave science with stochastic backgrounds:

First, how does the information contained in the stochastic signal compare to what we learn from resolvable binaries in the nearby Universe?
In Sec.~\ref{BBHbackground}, we demonstrate that the stochastic signal is dominated by unresolvable sources between redshifts $z\approx0.1$ and $3.5$;
thus, observations of the stochastic background will probe a BBH population that is distinct from directly resolvable sources in the more local Universe.

Second, what astrophysics and cosmology can we explore using results from stochastic searches?
In Sec.~\ref{model_selection}, we find that, while second-generation gravitational wave detectors may successfully measure the amplitude of the stochastic background, it is difficult to further distinguish between different models for the binary black hole background.

Third, how does the presence of the expected binary black hole background affect our ability to measure other potentially interesting backgrounds arising, e.g., from cosmic strings \cite{Kibble_1973, Caldwell_Allen_1992}, the core collapse of population III stars \cite{Crocker_MandicEA_2015}, or phase transitions in the early Universe \cite{Starobinsky_1979, KosowskyEA_1992, Giblin_Thrane_2014, 2000PhR...331..283M}?
In Sec.~\ref{ALTbackgrounds}, we show that the BBH background acts as a limiting foreground, significantly decreasing our sensitivity to other backgrounds of interest.

\section{INFORMATION CONTAINED IN THE BBH BACKGROUND} \label{BBHbackground}

A stochastic background of gravitational waves introduces a correlated signal in networks of terrestrial detectors.
Although this signal is much weaker than the detector noise, it is detectable by cross-correlating the strain data from two or more detectors.
For a two-detector network and an isotropic, unpolarized, and stationary Gaussian background, the optimal SNR of a cross-correlation search is given by \cite{Allen_Romano_1999}
	\begin{equation}
	\text{SNR} = \frac{3H_0 ^2}{10\pi^2}\sqrt{2T}\left[\intHalf \frac{\gamma^2(f) \Omega^2\gw(f)}{f^6 P_1(f)P_2(f)}\,df\right]^{1/2},
	\label{SNR}
	\end{equation}
where $P_i(f)$ is the noise power spectral density (PSD) of detector $i$, $\gamma(f)$ is the normalized isotropic overlap reduction function \cite{1992PhRvD..46.5250C}, and $T$ is the total accumulated coincident observation time. The energy density spectrum $\Omega\gw(f)$ of the stochastic background is defined as
	\begin{equation}
	\label{Omega_gw}
	\Omega\gw(f) = \frac{1}{\rho_c}\frac{d\rho\gw}{d \ln f},
	\end{equation}	
where $d\rho\gw$ is the energy density in gravitational waves per logarithmic frequency interval $d\ln f$ and $\rho_c = 3 H_0^2 c^2/8\pi G$ is the critical energy density required to close the Universe.
Here, $G$ is Newton's constant, $c$ is the speed of light, and $H_0$ is the Hubble constant.
We assume a standard ``737 cosmology,'' with $H_0=70 \,\text{km}\,\text{s}^{-1}\,\text{Mpc}^{-1}$, $\Omega_m = 0.3$, and $\Omega_\Lambda=0.7$.

\begin{figure}
\centering
\includegraphics[width=.48\textwidth]{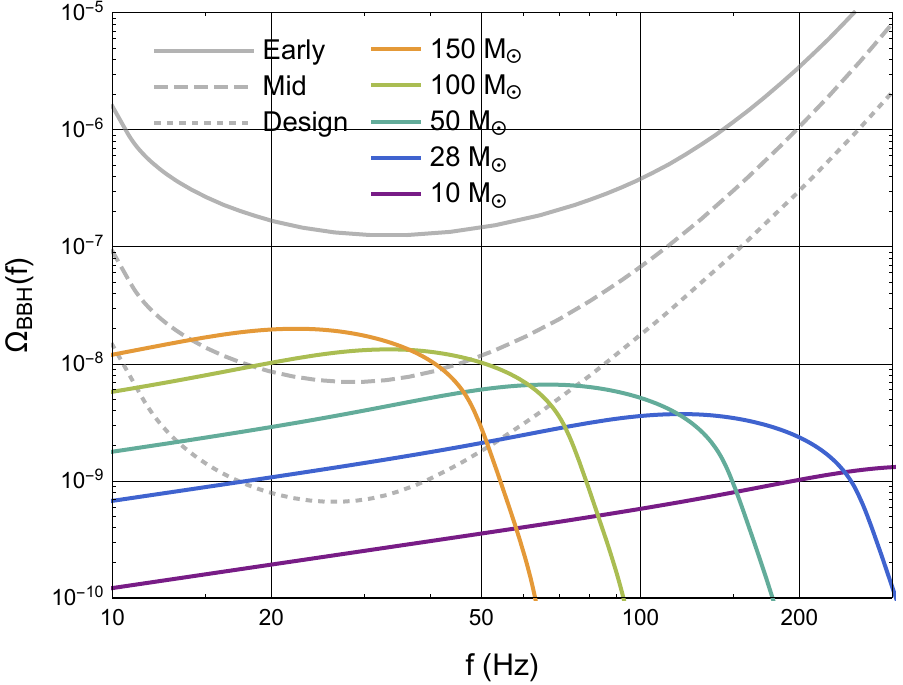}
\caption{
Binary black hole backgrounds of various chirp masses, assuming a local coalescence rate of $R_0 = 16\,\text{Gpc}^{-3}\text{yr}^{-1}$ and the \texttt{Fiducial} model for the stochastic background.
Power-law integrated curves \cite{Thrane2013a} for one year of integration with Advanced LIGO at early, middle, and design sensitivity are shown for comparison.
Approximately 95\% of the signal-to-noise ratio comes from a band spanning $\unit[15-45]{Hz}$.
The shape and amplitude of $\Omega\bbh(f)$ depend on the average chirp mass of the BBH population.
As $\mathcal{M}_c$ increases with fixed $R_0$, the peak value of $\Omega\bbh(f)$ grows like $\mathcal{M}_c^{5/3}$, while the knee frequency $f_\text{max}$ at which the peak occurs scales as $f_\text{max} \sim 1/\mathcal{M}_c$.}
\label{fig:MandOmega}
\end{figure}

The energy density spectrum of a binary black hole background is determined in part by the binary chirp mass $\mathcal{M}_c=\eta^{3/5} M$, where $M$ is the binary's total mass and $\eta$ its symmetric mass ratio.
Figure~\ref{fig:MandOmega} shows example energy density spectra for stochastic BBH backgrounds of various ``average chirp masses,"
[more precisely, the background depends on the average $\overline{\mathcal{M}_c^{5/3}}$;
hereafter, the average chirp mass $\mathcal{M}_c$ refers to $(\overline{\mathcal{M}_c^{5/3}})^{3/5}$],
assuming equal mass binaries with $\eta = 0.25$.
Also shown in Fig.~\ref{fig:MandOmega} are power-law integrated (PI) curves \cite{Thrane2013a} indicating the sensitivity of the stochastic search after one year of integration with Advanced LIGO at early, middle, and design sensitivity.
Power-law integrated curves are defined such that a power-law energy density spectrum drawn tangent to the PI curve will give $\text{SNR} = 1$ after one year. 
More generally, energy density spectra lying above a PI curve have $\text{SNR}\gtrsim1$ after one year, while those below have $\text{SNR}\lesssim1$.

We adopt the \texttt{Fiducial} model of Ref.~\cite{gw150914_stoch}, with BBH energy density spectra given by \cite{Phinney2001,2011ApJ...739...86Z,Wu_EA_2012,gw150914_stoch}
	\begin{equation}
	\label{Omega_f}
	\Omega\bbh(f) = \frac{f}{\rho_c}
		\int \frac{\frac{dE\bbh [f(1+z)]}{df} R_m(z)}{(1+z)H(z)}\,dz,
	\end{equation}
where $dE\bbh/df$ is the source-frame energy spectrum of a single BBH source~\cite{Ajith_EA_2008} (see Appendix~\ref{bbhappendix}).
Since energy and frequency are identically redshifted, $dE\bbh/df$ is in fact redshift invariant, depending on $z$ only through its argument as shown in Eq.~\eqref{Omega_f}.
$H(z) = H_0\sqrt{\Omega_m(1+z)^3+\Omega_\Lambda}$ describes the evolution of the Hubble parameter in a flat universe.
Finally, $R_m(z)$ is the BBH merger rate per comoving volume as measured in the source frame;
the factor of $(1+z)$ in the denominator of Eq.~\eqref{Omega_f} converts this rate into the detector frame.

We assume that $R_m(z)$ traces the star formation rate $R_*(z)$, subject to a time delay $t_d$ between a binary's formation and merger \cite{gw150914_stoch}:
	\begin{equation}
	\label{rate}
	R_m(z) = R_0 \frac{\int_{t_\text{min}}^{t_\text{max}} {R_\ast}[z_f(t_d,z)] \, F[z_f(t_d,z)] \, P(t_d)dt_d}
		{\int_{t_\text{min}}^{t_\text{max}} {R_\ast}[z_f(t_d,0)] \, F[z_f(t_d,0)] \, P(t_d)dt_d}.
	\end{equation}
Here, $R_0$ is the local coalescence rate at $z=0$, $P(t_d)$ is the probability of a time delay $t_d$, and $z_f(t_d,z)$ is the formation redshift corresponding to merger at redshift $z$.
We take $t_\text{min}=50$ Myr to be the minimum time required for binary evolution through merger, and integrate up to $t_\text{max} = 13.5$ Gyr.
We assume $P(t_d)\propto t_d^{-1}$ for $t_d \geq t_\text{min}$~\cite{2012ApJ...759...52D}.
For $R_*(z)$, we adopt the star formation rate presented in Sec. 2.1 of Ref.~\cite{2015MNRAS.447.2575V}, based on the observed gamma-ray burst rate~\cite{Kistler2013}.
We also assume that binary black holes are born preferentially in low-metallicity environments, multiplying $R_*(z)$ by the fraction  $F(z)$ of stars formed with metallicities $Z < Z_\odot/2$~\cite{gw150914_stoch}, where $Z_\odot = 0.02$ is the solar metallicity; see Appendix~\ref{sfrappendix} for details.
Below, we also consider the \texttt{LowMetallicity} model of Ref.~\cite{gw150914_stoch}, which instead assumes progenitor metallicities $Z<Z_\odot/10$.

\begin{figure*}
\centering
    	\includegraphics[width=.44\textwidth]{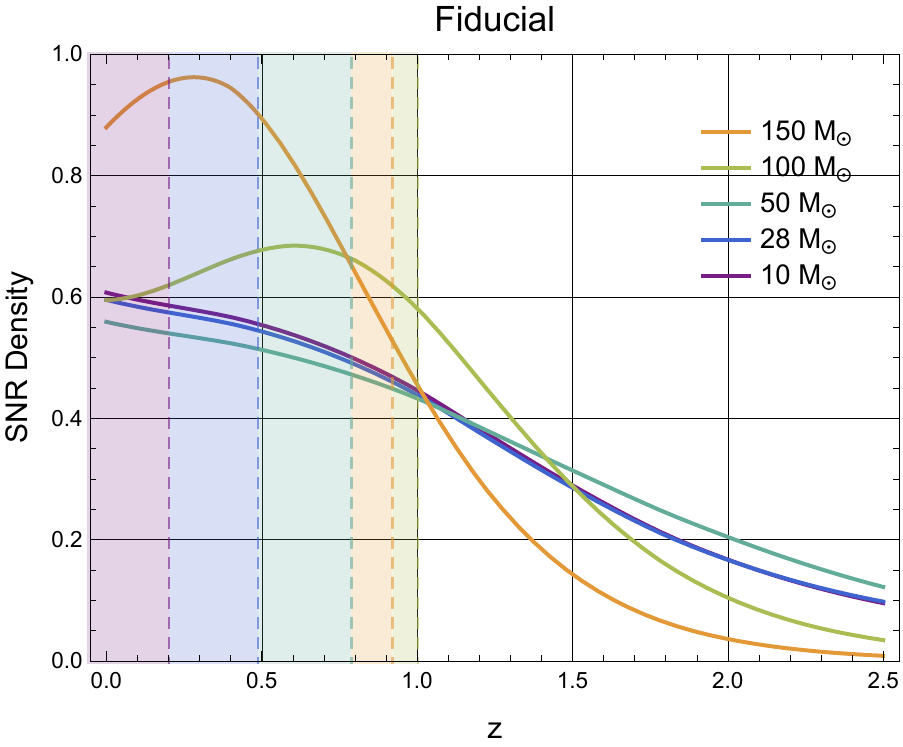} \hspace{5mm}
  	\includegraphics[width=.44\textwidth]{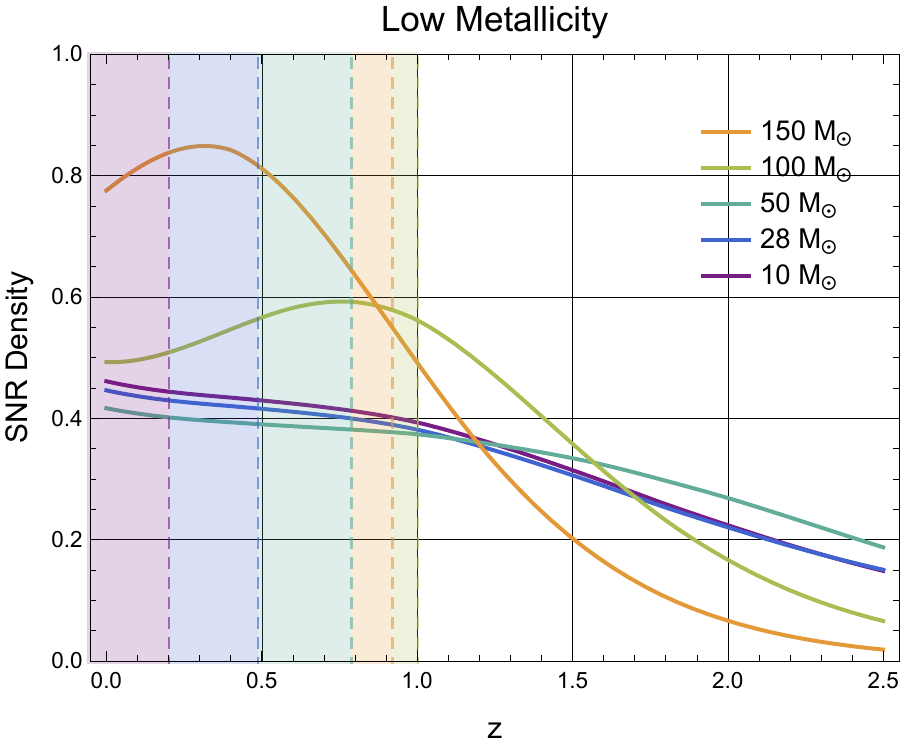} \\
  	\includegraphics[width=.44\textwidth]{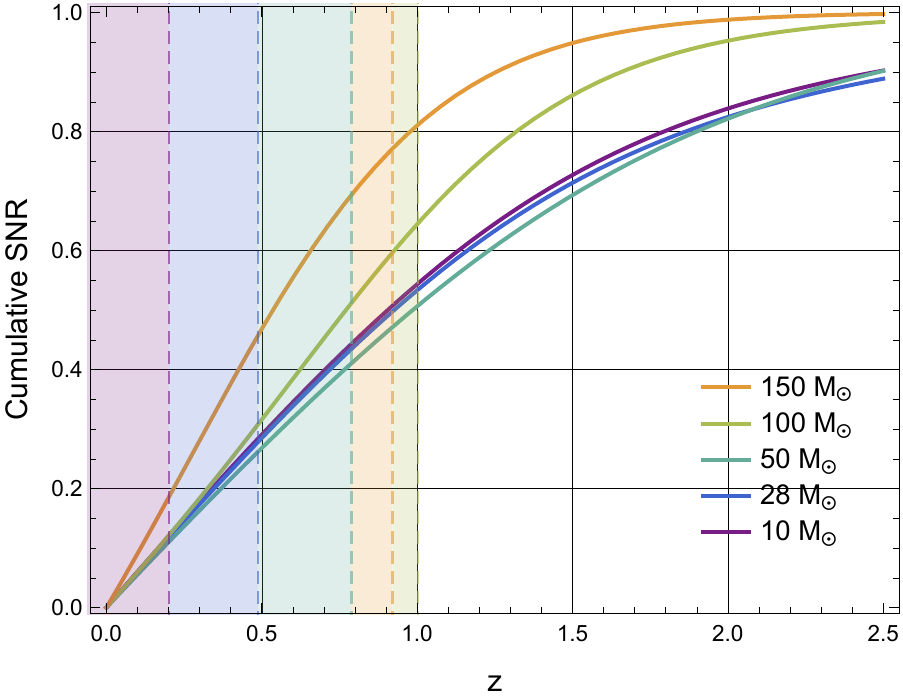} \hspace{5mm}
  	\includegraphics[width=.44\textwidth]{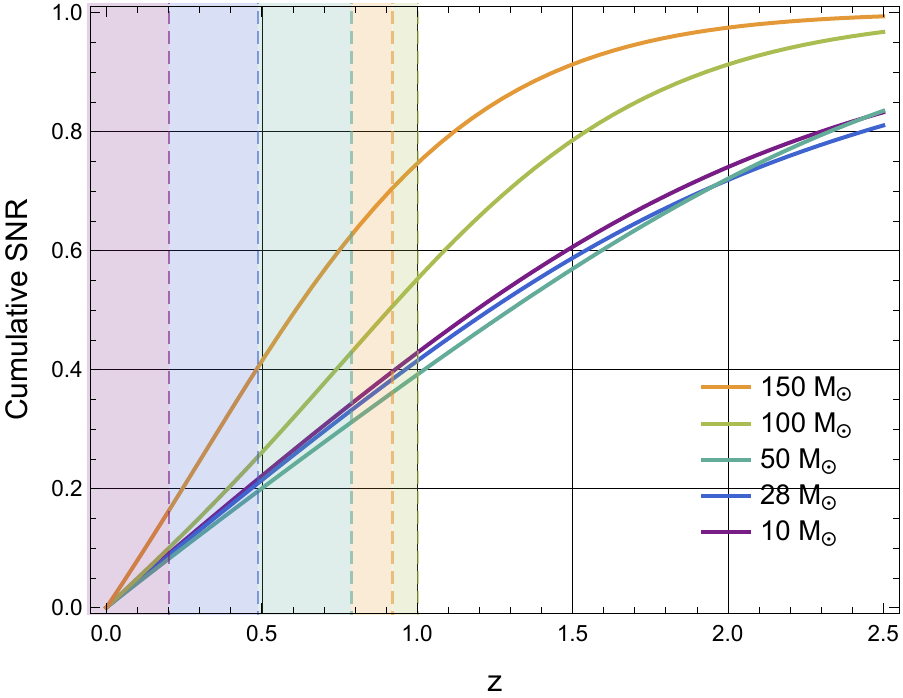}
  \caption{
    \textit{Top}: SNR density $d(\text{SNR})/dz$ for various choices of chirp mass, assuming the \texttt{Fiducial} (\textit{left}) and \texttt{LowMetallicity} (\textit{right}) background models.
    Each curve is normalized such that its integral over all redshifts is 1.
    \textit{Bottom}: Cumulative SNR, found by integrating SNR density from $(0,z)$ assuming the \texttt{Fiducial} (\textit{left}) and 
    \texttt{LowMetallicity} (\textit{right}) models.
    For each choice of mass, the total SNR is normalized to 1.
    The dashed vertical lines indicate Advanced LIGO's ``threshold redshifts'' $z_{50\%}$ for binaries of each chirp mass (given by the respective color).
    More than 50\% of binaries merging at $z<z_{50\%}$ (in the shaded regions) will be individually resolvable.
    Most binaries at $z>z_{50\%}$ cannot be individually resolved, but contribute to the measured stochastic signal.
    Note that, because much of the signal from high-redshift $\mathcal{M}_c = 150 M_\odot$ binaries is redshifted out of Advanced LIGO's sensitivity band, $z_{50\%}$ for such binaries is \textit{lower} than for those with $\mathcal{M}_c = 100M_\odot$.
    }
  \label{fig:dSNR}
\end{figure*}

Unlike direct searches for binary coalescences, the results of which are dominated by the closest sources, the stochastic background is dominated by distant sources.
To explain this simply, we imagine an idealized static Universe with a constant merger rate and no cosmological expansion.
The gravitational wave energy density $d\Omega$ contributed by binaries within a thin spherical shell of radius $r$ and thickness $dr$ scales like $d\Omega \sim h^2 dN \sim 1/r^2 dN$, where $h$ is gravitational wave strain and $dN$ is the number of sources within the shell.
In our idealized universe, BBH binaries are equally distributed in volume, so $dN \sim r^2 dr$, giving $d\Omega\sim dr$. 
The background contribution from any shell is therefore constant, independent of distance.
Since the number of such shells beyond Advanced LIGO's horizon distance is much greater than the number within, the stochastic background is dominated by distant, unresolvable sources.
(This is a reformulation of Olber's paradox.)

In reality, the BBH population is not uniformly distributed in volume; we assume it traces the star formation rate via Eqs.~\eqref{Omega_f} and~\eqref{rate}.
In order to more rigorously investigate the SNR contribution from binaries at different redshifts, we define the ``SNR density''
\begin{equation}
  \frac{d(\text{SNR})}{dz} = \frac{2T}{\text{SNR}} \left(\frac{3H_0 ^2}{10\pi^2}\right)^2 \intHalf \frac{\gamma^2(f) \Omega\bbh(f) \frac{d\Omega\bbh}{dz}(f,z)}{f^6 P_1(f)P_2(f)}df, 
\end{equation}
with
\begin{equation}
  \frac{d\Omega\bbh}{dz}(f,z) = \frac{f}{\rho_c} \frac{ \frac{dE\bbh[f(1+z)]}{df} R_m(z)}{(1+z)H(z)}.
\end{equation}

SNR density for design-sensitivity Advanced LIGO is plotted as a function of $z$ in Fig.~\ref{fig:dSNR} for several choices of chirp mass, assuming the \texttt{Fiducial} and \texttt{LowMetallicity} models.
Also shown are the cumulative SNRs obtained by integrating the SNR density up to some cutoff $z$.
For purposes of comparison, the curves shown are each normalized to total $\text{SNR}=1$.
In each figure, the dashed vertical lines indicate ``threshold redshifts'' $z_{50\%}$ beyond which BBHs of each chirp mass (indicated by the respective colors) are individually resolvable less than 50\% of the time (see Appendix~\ref{rangeappendix} for details).
The redshifts $z_{50\%}$ therefore indicate the typical range of a direct search for compact binary coalescences -- binaries at redshifts $z<z_{50\%}$ are, on average, directly resolvable, while those at $z>z_{50\%}$ are not.

For binaries like GW150914, with $\mathcal{M}_c \approx 28M_\odot$ and $z_{50\%}\approx0.5$, approximately $70\%$ of the stochastic SNR is due to unresolvable binaries when assuming the \texttt{Fiducial} model.
In this case, $90\%$ percent of the stochastic signal is contributed by sources between $z\approx 0.1$ and $3.5$, and $50\%$ is due to binaries beyond $z\approx0.9$.
These precise values depend on the specific choice of background model.
The \texttt{LowMetallicity} model, for instance, predicts that $80\%$ of the SNR is due to unresolvable sources, with $90\%$ percent of the signal contributed by binaries between $z\approx0.1$ and $4.2$.
For very high-mass systems ($\mathcal{M}_c=150 M_\odot$), $z_{50\%}\approx0.9$, and so a much larger fraction of the stochastic SNR is due to resolvable sources.
In this case, only approximately $20\%$ of the stochastic signal remains due to unresolvable binaries.

It is interesting to see how SNR density changes with average chirp mass.
For $\mathcal{M}_c \lesssim 50\,M_\odot$, the curves are all similar because the knee frequency of $\Omega\bbh(f)$ is outside the sensitive part of the band; 95\% of the $\text{SNR}$ is contained between $\approx15$ and $45\,\text{Hz}$ (see the PI curves in Fig.~\ref{fig:MandOmega}).
At $\mathcal{M}_c \approx 100\,M_\odot$, the SNR density distribution shifts to higher $z$ because the loud merger signal from high-$z$ sources is redshifted into the most sensitive band.
Finally, as $\mathcal{M}_c$ increases further to $\gtrsim 150 M_\odot$, the merger signal from high-$z$ signals begins to leave the observing band entirely, leaving mostly signal from low-$z$ sources.

\section{STOCHASTIC MODEL SELECTION} \label{model_selection}

Valuable astrophysical information is contained in the BBH background, including the masses and merger rates of distant BBH populations inaccessible to direct searches for compact binary coalescences.
The degree to which this information can be extracted, however, depends on our ability to perform model selection and parameter estimation.
Model selection and parameter estimation has been shown to be difficult for astrophysical backgrounds dominated by low-mass binaries of several $M_\odot$~\cite{paramest}, which only depart from $\Omega(f)\propto f^{2/3}$ power laws at frequencies above $\sim\unit[1]{kHz}$.
The low stochastic search sensitivity above $50$ Hz suggests that this high-frequency behavior will be extremely difficult to observe.

Backgrounds of more massive BBHs are shifted to lower frequencies (see, e.g., Fig.~\ref{fig:MandOmega}), where non-power-law spectral features are increasingly visible to ground-based detectors.
This suggests that black hole backgrounds may be more promising targets for model selection and parameter estimation.
In order to evaluate the prospects for model selection on BBH backgrounds, we investigate at what point an astrophysical \texttt{Fiducial} background can be distinguished from a power-law spectrum:
\begin{equation}
\label{powerlaw}
\Omega\plaw(f) = \Omega_0 \left(\frac{f}{f_0}\right)^{2/3},
\end{equation}
where $f_0$ is an arbitrary reference frequency.

The standard stochastic search employs a cross-correlation statistic $\hat Y(f) \propto \tilde s^*_1(f)  \tilde s_2(f)$ that is proportional to the strain cross power between the signals $\tilde s_1$ and $\tilde s_2$ measured by two detectors~\cite{Allen_Romano_1999}.
The expectation value and variance of the cross-correlation statistic $\hat Y(f)$ in a frequency bin of width $df$ are, with appropriate normalization,
	\begin{equation}
	\langle \hat Y(f) \rangle = \Omega(f)
	\end{equation}
and
	\begin{equation}
	\sigma^2(f) = \frac{1}{2Tdf} \left( \frac{10\pi^2}{3H_0^2}\right)^2 \frac{f^6 P_1(f) P_2(f)}{\gamma(f)^2}.
	\end{equation}
Here, $\Omega(f)$ is the true gravitational wave background.
When adopting a particular model $\Omega_M(f)$ for the stochastic background, the likelihood for $\hat Y(f)$ is the Gaussian~\cite{Allen_Romano_1999,paramest}
	\begin{equation}
	\label{singleLikelihood}
	\mathcal{L}_f (\hat Y\,|\,\Omega_M) \propto \exp \left( - \frac{ [\hat Y(f) - \Omega_M(f)]^2}{2 \sigma^2(f)} \right).
	\end{equation}
The value $\hat Y(f)$ measured in any single experiment is a random variable, depending on the particular noise instantiation $\delta\Omega(f)$ through $\hat Y(f) = \Omega(f) + \delta\Omega(f)$.
The noise $\delta\Omega(f)$ is itself Gaussian distributed about zero with variance $\sigma^2(f)$.
In the absence of real data, we cannot compute Eq.~\eqref{singleLikelihood}, but can instead calculate the ensemble-averaged likelihood
	\begin{equation}
	\langle \mathcal{L}_f (\hat Y\,|\,\Omega_M)\rangle  \propto \exp \left( - \frac{ [\Omega(f) - \Omega_M(f)]^2}{4 \sigma^2(f)} \right),
	\end{equation}
obtained by marginalizing over $\delta\Omega(f)$;
this result is similar to Eq.~\eqref{singleLikelihood}, but with an additional factor of $1/2$ in the exponential.
Simply assuming $\delta \Omega = 0$ produces an overly optimistic estimate of an experimental likelihood.
The full (ensemble-averaged) likelihood is the product $\mathcal{L} \propto \prod_f \langle \mathcal{L}_f \rangle$, given by
	\begin{equation}
	\label{L3}
	\mathcal{L}(\Omega\,|\,\Omega_M) = \mathcal{N} \exp \left[ -\frac{1}{4} \left(\Omega-\Omega_M \,|\, \Omega-\Omega_M\right) \right],
	\end{equation}
where $\mathcal{N}$ is a normalization factor and we have defined the inner product
	\begin{equation}
	\label{innerproduct}
	\left(A\,|\,B\right) = 2 T \left(\frac{3 H_0^2}{10\pi^2}\right)^2  \intHalf \frac{\gamma(f)^2 \tilde A(f) \tilde B(f)}{f^6 P_1(f) P_2(f)}df.
	\end{equation}
Note that $\Omega$, not $\hat Y$, appears on the left-hand side of Eq.~\eqref{L3}, since this ensemble-averaged likelihood depends only on the expectation value $\langle \hat Y(f)\rangle = \Omega(f)$.

Given an underlying BBH background described by the \texttt{Fiducial} model, we investigate the maximum likelihood ratio $\mathcal{R}=\mathcal{L}\ml(\Omega\bbh\,|\,\Omega\bbh)/\mathcal{L}\ml(\Omega\bbh\,|\,\Omega\plaw)$ between the \texttt{Fiducial} and power-law models.
Large values of $\mathcal{R}$ indicate that the \texttt{Fiducial} model is (correctly) preferenced over the power-law background model; values close to $\mathcal{R}=1$ indicate that the two models are indistinguishable.
The maximum likelihood when correctly assuming the \texttt{Fiducial} model is $\mathcal{L}\ml(\Omega\bbh\,|\,\Omega\bbh) = \mathcal{N}$,
since the background itself is contained within the space of BBH models.
The maximum likelihood when incorrectly assuming a power-law model can be derived analytically.
The power-law model has one free parameter -- the amplitude $\Omega_0$.
The amplitude maximizing the likelihood Eq. \eqref{L3} satisfies $d \mathcal{L}(\Omega\bbh\,|\,\Omega\plaw)/d\Omega_0 = 0$, which is solved to give
	\begin{equation}
	\label{maxAmp}
	\Omega_0\ml = \frac{ \left(\omega \,|\, \Omega\bbh \right)}{\left(\omega\,|\,\omega\right)}
	\end{equation}
where $\omega(f) = (f/f_0)^{2/3}$.
The corresponding maximum likelihood for the power-law model is
	\begin{equation}
	\label{maxLikelihoodPL}
	\mathcal{L}\ml(\Omega\bbh\,|\,\Omega\plaw) = \mathcal{N} \exp \left\{ -\frac{1}{4}
		\left( \left(\Omega\bbh\,|\,\Omega\bbh\right) - \frac{ \left(\omega\,|\,\Omega\bbh\right)^2}{\left(\omega\,|\,\omega\right)}\right)\right\}.
	\end{equation}
	
\begin{figure*}
	\includegraphics[width=.4\textwidth]{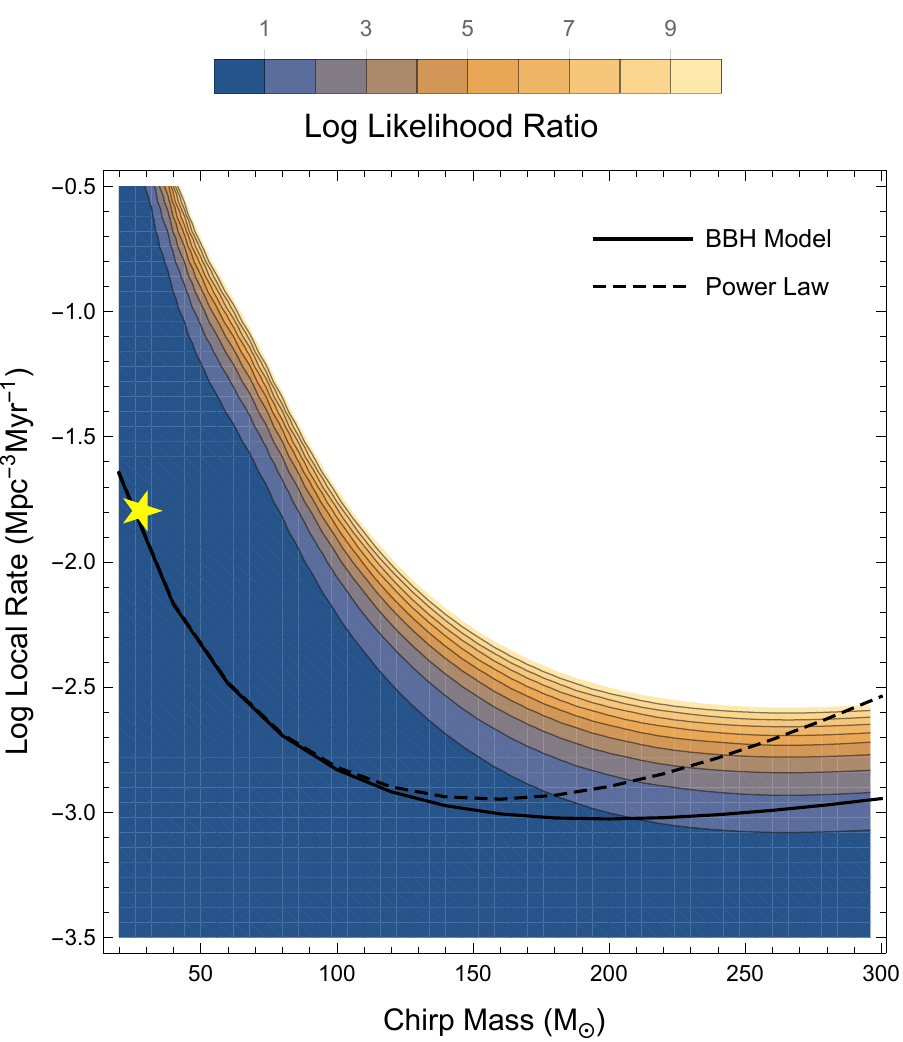} \hspace{5mm}
	\includegraphics[width=.4\textwidth]{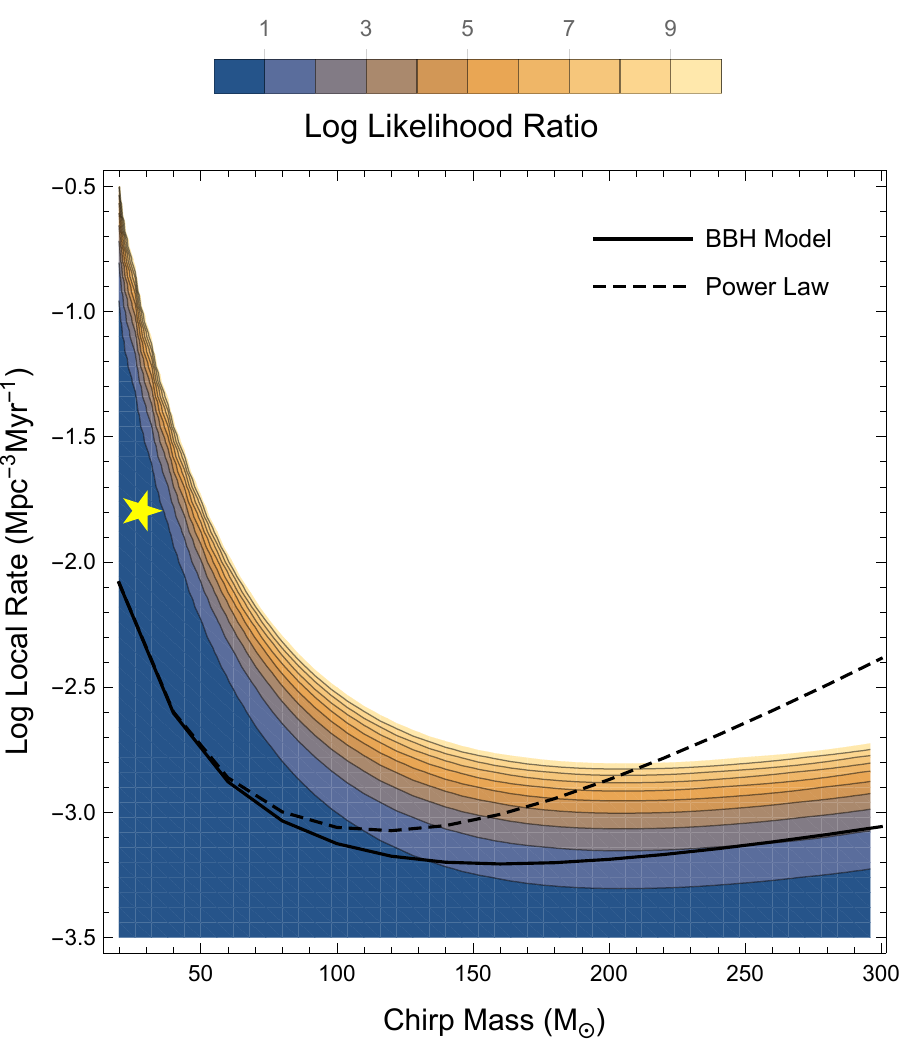}
\caption{
Contours of maximum log-likelihood ratios $\ln\mathcal{R}$ between the \texttt{Fiducial} and power-law background models [Eqs.~\eqref{Omega_f} and \eqref{powerlaw}, respectively] for Advanced LIGO (H1-L1) (\textit{left}) and colocated detectors (H1-H2) (\textit{right}), as a function of the background's average chirp mass and local coalescence rate.
The results shown assume three years of integration time at design sensitivity.
The solid and dashed black curves indicate the local coalescence rates at which a BBH background is detectable with $\mathrm{SNR}=3$ after three years when assuming the \texttt{Fiducial} and power-law models, respectively, and the star indicates the background associated with GW150914~\cite{gw150914_stoch}.
Although the background inferred from GW150914 may be marginally detectable with Advanced LIGO after three years of observation, it is indistinguishable from a simple power-law model.
The background remains indistinguishable from a power law even for colocated detectors, which are predicted to make a strong detection of the BBH background.
}
\label{likelihoodContoursMassRate}
\end{figure*}

When formally comparing models with different numbers of parameters (such as the one-parameter power law versus the many-parameter \texttt{Fiducial} model), one could alternatively calculate a Bayes factor rather than a maximum likelihood ratio.
However, the Bayes factor is approximated by the maximum likelihood ratio, multiplied by an additional ``Occam's factor'' penalizing the more complex of the two models~\cite{Cornish_Romano_2015}.
The inclusion of the Occam's factor here would only serve to penalize the \texttt{Fiducial} model;
by neglecting it here, we are showing the most optimistic prospects for discerning the form of an astrophysical BBH background.
Additionally, when model parameters are not informative, the associated Occam's factor is near unity and the maximum likelihood ratio well approximates the Bayes factor.

Figure~\ref{likelihoodContoursMassRate} shows contours of the maximum log-likelihood ratio $\ln\mathcal{R}$ as a function of the local coalescence rate and chirp mass after three years of observation at design sensitivity.
The solid black curve indicates the rates above which a BBH background is detectable with optimal $\mathrm{SNR}=3$ after three years when correctly assuming the \texttt{Fiducial} model.
The dashed black curve similarly indicates rates above which BBH backgrounds are detectable with $\mathrm{SNR}=3$ when assuming a power-law model.
This is \textit{not} the optimal SNR, since the space of power-law models does not contain the true BBH spectrum. 
The best-fit background parameters inferred from GW150914 \cite{gw150914_rate,gw150914_pe,gw150914_stoch} are indicated by a star.
Over a large region of parameter space, $\ln\mathcal{R}\lesssim 1$;
in this region, the power-law and \texttt{Fiducial} models cannot be distinguished.
Only for chirp masses and local rates much larger than those implied by GW150914 is $\ln\mathcal{R}>1$.
While Advanced LIGO is therefore likely to detect the stochastic background associated with GW150914, such a background is indistinguishable from a simple power law.
In particular, $\approx6000$ years of observation at design sensitivity are required to attain $\ln\mathcal{R}=3$.

The Advanced LIGO network consists of two interferometers at Hanford, Washington and Livingston, Louisiana.
The sensitivity of the Hanford-Livingston (H1-L1) network to a BBH background is ultimately limited at high frequencies by the overlap reduction function $\gamma(f)$, which rapidly approaches zero for $f\gtrsim 60$ Hz~\cite{1992PhRvD..46.5250C}.
During Initial LIGO, a third interferometer (H2) was present at Hanford, colocated and co-oriented with H1~\cite{H1H2_2014}.
With a constant overlap reduction function of $\gamma_\text{H1-H2}(f) = 1$, the H1-H2 pair is significantly more sensitive at high frequencies than H1-L1.
While there are currently no plans to reinstall a second interferometer at Hanford during Advanced LIGO, it is interesting to consider the performance of a hypothetical H1-H2 network of colocated 4-km aLIGO interferometers.
Fig.~\ref{likelihoodContoursMassRate} also shows maximum likelihood ratios between the \texttt{Fiducial} and power law models for this hypothetical H1-H2 network.
Although the BBH background is detectable by the H1-H2 network after three years, it remains indistinguishable from a power law. 
Approximately 50 years of observation with design-sensitivity colocated detectors are required to reach $\ln\mathcal{R}=3$.
Although this represents a factor $\approx120$ improvement over the H1-L1 performance above, it is nevertheless an impractically long time.

\begin{figure*}
	\includegraphics[width=.4\textwidth]{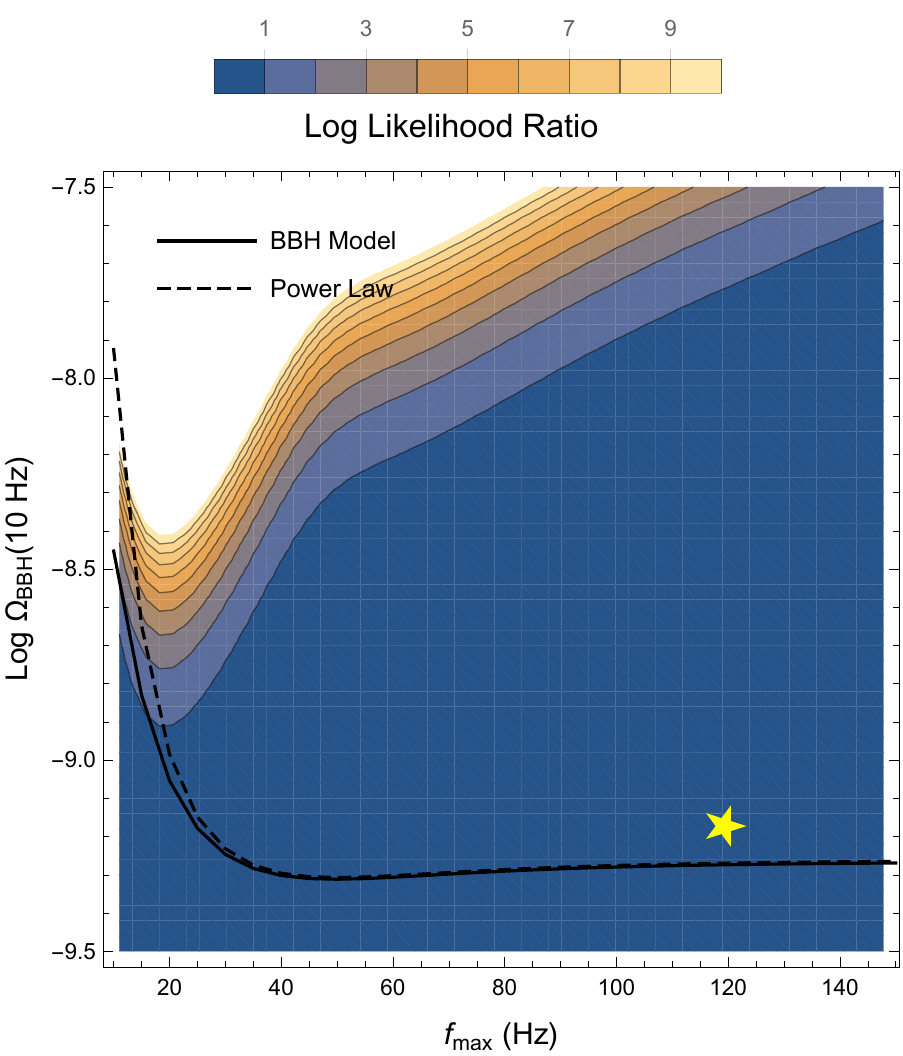} \hspace{5mm}
	\includegraphics[width=.4\textwidth]{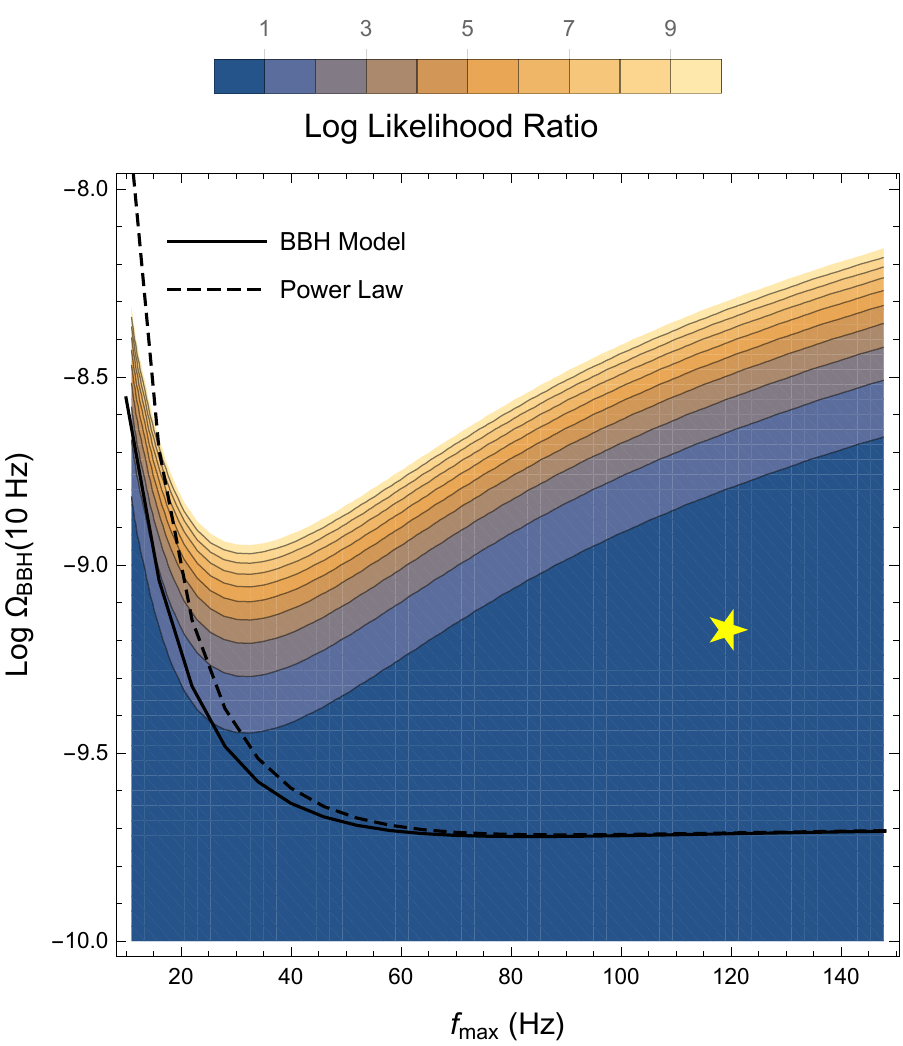}
\caption{
Maximum log-likelihood contours between the astrophysical and power-law models, as a function $f_\text{max}$ (see Fig.~\ref{fig:MandOmega}) and the background's amplitude at 10 Hz.
Results are shown for Advanced LIGO (\textit{left}) and a network of colocated aLIGO detectors (\textit{right}), assuming three years of integration at design sensitivity.
Advanced LIGO is best able to distinguish realistic background models from power laws for frequencies $f_\text{max}$ between 10 and 50 Hz, corresponding to the most sensitive frequency band for the stochastic search.
As in Fig.~\ref{likelihoodContoursMassRate}, solid and dashed black curves show the amplitudes at which a background is detectable with $\text{SNR}=3$ after three years, using the \texttt{Fiducial} and power-law models, respectively.
The star indicates the \texttt{Fiducial} background associated with GW150914.}
\label{likelihoodContoursFreqAmp}
\end{figure*}

In Fig.~\ref{likelihoodContoursFreqAmp}, contours of $\ln\mathcal{R}$ are instead shown in terms of the background amplitude at 10 Hz (which scales as $\Omega\sim \mathcal{M}_c^{5/3}R_0$) and the frequency $f_\text{max}$ at which the background's energy density is at a maximum ($f_\text{max} \sim 1/\mathcal{M}_c$; see Fig.~\ref{fig:MandOmega}).
From this figure, it is apparent that the only backgrounds distinguishable from power laws using H1-L1 are those for which $f_\text{max}\sim10-50$ Hz, which corresponds to the most sensitive frequency band for the isotropic stochastic search.
The H1-H2 network shows sensitivity across a broader frequency band, as this configuration avoids the penalty associated with the overlap reduction function at high frequencies.

Any configuration of advanced detectors appears unlikely to differentiate an astrophysical BBH model from a simple power law.
Hence, parameter estimation and model selection on the BBH background is limited to studying only its amplitude rather than its shape, and efforts to simultaneously constrain multiple parameters (e.g., $\mathcal{M}_c$ and $R_0$) from the stochastic background alone will be thwarted by large degeneracies.
Some sensitivity can be gained, however, by applying direct CBC measurements as priors on stochastic background parameters.
With tight priors on the chirp mass and local rate, for instance, the stochastic search becomes increasingly sensitive to amplitude differences between different models of the BBH merger rate and redshift distribution.

\begin{figure}
\includegraphics[width=.48\textwidth]{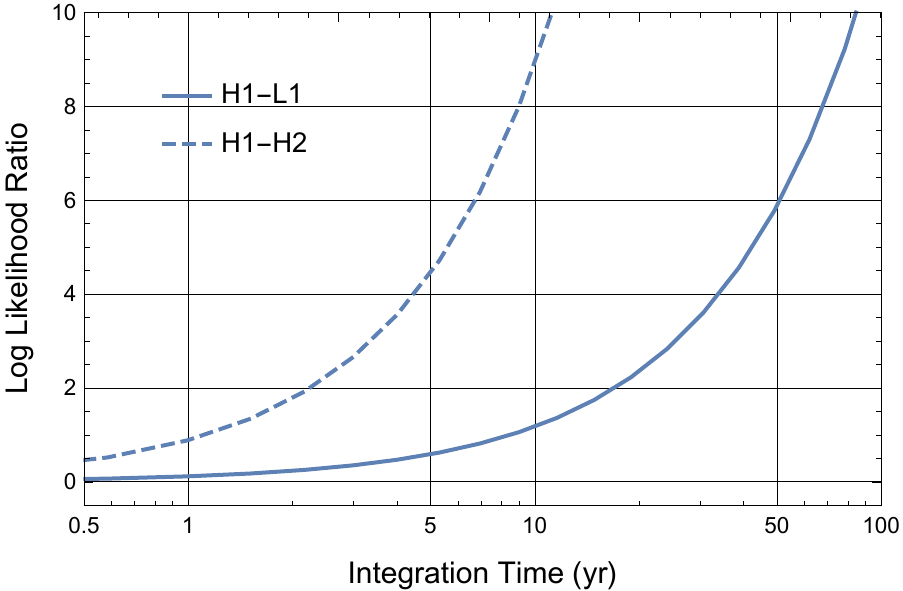}
\caption{
Projected log-likelihood ratios $\ln\mathcal{R}$ between the \texttt{Fiducial} and \texttt{LowMetallicity} background models, as a function of observation time with Advanced LIGO (H1-L1, solid curves).
We assume that the underlying BBH background is given by the \texttt{Fiducial} model, with chirp mass $\mathcal{M}_c=28M_\odot$ and local coalescence rate $R_0 = 16\text{\,Gpc}^{-3}\text{yr}^{-1}$ following GW150914.
We also include log-likelihoods for a network of colocated Advanced LIGO detectors (H1-H2, dashed curves).
}
\label{fig:sfrlikelihoods}
\end{figure}

For a GW150914-like background, for instance, Fig.~\ref{fig:sfrlikelihoods} shows likelihood ratios between the \texttt{Fiducial} and \texttt{LowMetallicity} models as a function of observation time, for both the H1-L1 and H1-H2 detector networks (solid and dashed curves, respectively).
We take the \texttt{Fiducial} model as the ``true'' background, and assume delta-function priors on the average chirp mass and local coalescence rate.
Even in this most optimistic case, at least 25 years of observation with H1-L1 are required to distinguish (with log-likelihood ratio $\ln\mathcal{R}=3$) between these models.
Colocated detectors, however, begin to distinguish between the \texttt{Fiducial} and \texttt{LowMetallicity} models in only three years.
In a more careful treatment using realistic priors on the average chirp mass and local coalescence rate, we find that approximately 30 years of observation are required to distinguish between background models using H1-L1, while 10 years are required with H1-H2.

As a rule of thumb, it is only possible to distinguish between two astrophysical scenarios if the difference $\Delta\Omega(f)$ between their predicted spectra exceeds the sensitivity of the detector network.
For the Advanced LIGO Hanford-Livingston network operating at design sensitivity, a deviation of 
\begin{equation}
  \Delta\Omega(f) \gtrsim 10^{-9} (f/\unit[10]{Hz})^{2/3}
  (\unit[1]{yr}/T)^{1/2} 
\end{equation}
is required in order to select between two models with $3\sigma$ significance.

The above analysis assumes a standard cross-correlation pipeline, optimal for a stochastic background that is stationary, isotropic, and Gaussian. However, the BBH stochastic background is non-Gaussian~\cite{gw150914_stoch}, and it may be possible to improve upon the above results with future pipelines optimized for non-Gaussian backgrounds~\cite{popcorn,lionel,Cornish_Romano_2015}.

\section{RESOLVING ADDITIONAL BACKGROUNDS}\label{ALTbackgrounds}

Once a stochastic signal is observed by advanced detectors, a natural question will be:
is it consistent with the expected background from binary black hole mergers (the ``BBH-only'' hypothesis), or is there a contribution from something else, e.g. cosmic strings or cosmological sources (the ``BBH+'' hypothesis)?
In this sense, the BBH background now becomes a limiting foreground, obscuring the presence of additional, weaker background components.
As a simple scenario, consider the combined signal $\Omega(f) = \Omega\bbh(f) + \Omega_c$ from a \texttt{Fiducial} background of GW150914-like black holes (chirp mass $\mathcal{M}_c = 28 M_\odot$ and local rate $R_0 = 16\,\text{Gpc}^{-3}\text{Yr}^{-1}$) and a flat background $\Omega_c$ of cosmological origin.
How loud must the cosmological background be in order to be detectable against the BBH background?
This question is equivalent to asking: how loud must the stochastic signal be in order to detect a spectral index that is inconsistent with the BBH scenario?
Since we know that a potentially detectable background from BBHs is expected, thanks to the observation of GW150194, only observation of a spectral index inconsistent with $2/3$ can provide evidence of a distinct cosmological background.

This question can be cast as a model selection problem.
The simplest BBH-only model is an $f^{2/3}$ power law parametrized only by an amplitude $\Omega_0 \propto R_0 \mathcal{M}_c^{5/3}$:
\begin{equation}
\label{bbhOnly}
\Omega_\textsc{bbh--}(f) = \Omega_0 \left(\frac{f}{f_0}\right)^{2/3}.
\end{equation}
This model is valid if we restrict our attention to the $\mathcal{M}_c \lesssim 150 M_\odot$ regime, where a power law is indistinguishable from a realistic background as demonstrated in Sec. \ref{model_selection}.
For the BBH+ model, assume a power law plus a constant $\Omega_2$:
\begin{equation}
\label{bbhPlus}
\Omega_\textsc{bbh+}(f) = \Omega_1 \left(\frac{f}{f_0}\right)^{2/3} + \Omega_2.
\end{equation}

As in Sec. \ref{model_selection}, we consider the maximum likelihood ratio $\mathcal{R}=\mathcal{L}\ml(\Omega\,|\,\Omega_\textsc{bbh+})/\mathcal{L}\ml(\Omega\,|\,\Omega_\textsc{bbh--})$ between these models, with likelihoods defined as in Eq.~\eqref{L3}.
The ``BBH-only'' likelihood is maximized by the amplitude $\Omega\ml_0$ given in Eq.~\eqref{maxAmp} [replacing $\Omega\bbh(f)$ with the combined background $\Omega(f)=\Omega\bbh(f) + \Omega_c$ considered here].
The ``BBH+'' likelihood is maximized by
\begin{equation}
\begin{aligned}
\Omega\ml_1 &= \frac{(\omega \,|\, 1)(\Omega\,|\,1) - (1\,|\,1)(\Omega\,|\,\omega)}
	{ (\omega\,|\,1)^2 - (\omega\,|\,\omega)(1\,|\,1)}, \\
\Omega\ml_2 &= \frac{(\omega \,|\, 1)(\Omega \,|\, \omega) - ( \omega\,|\,\omega)( \Omega \,|\, 1)}
	{ (\omega\,|\,1)^2 - (\omega\,|\,\omega)(1\,|\,1)}.
\end{aligned}
\end{equation}

\begin{figure*}
  \centering
  \includegraphics[width=.48\textwidth]{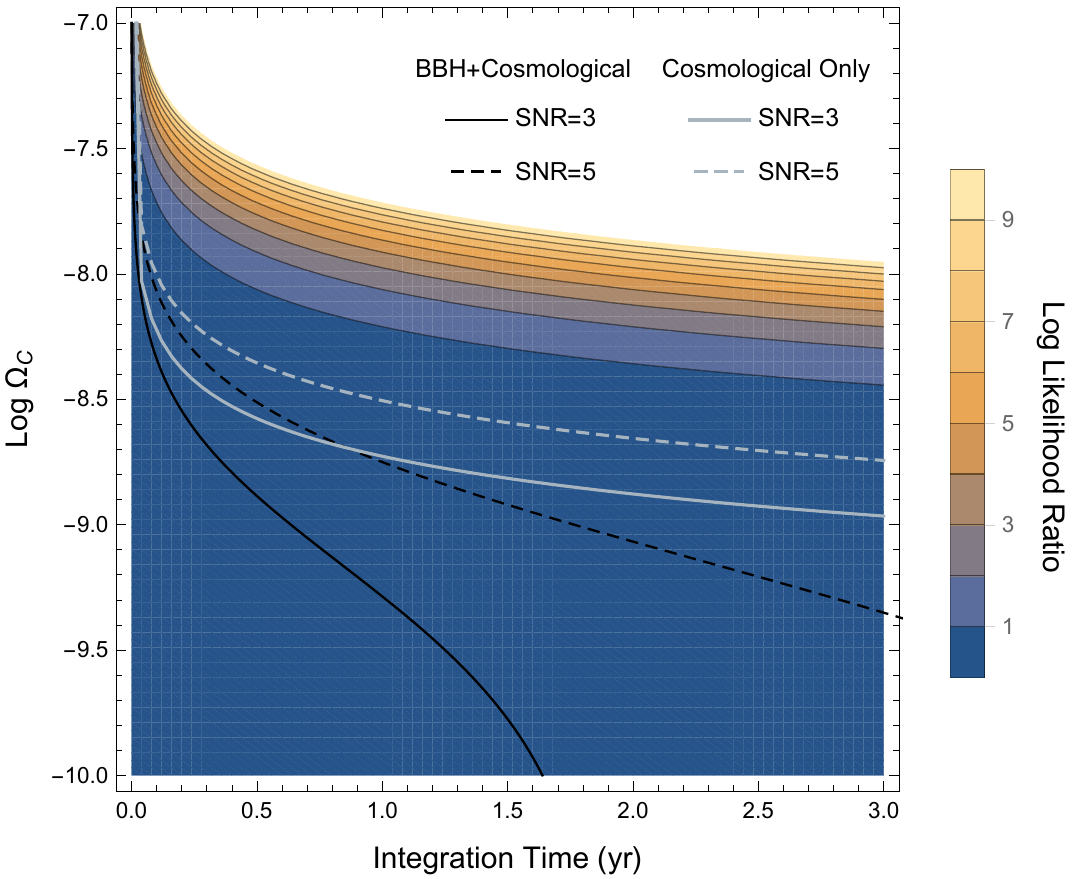}
  \includegraphics[width=.48\textwidth]{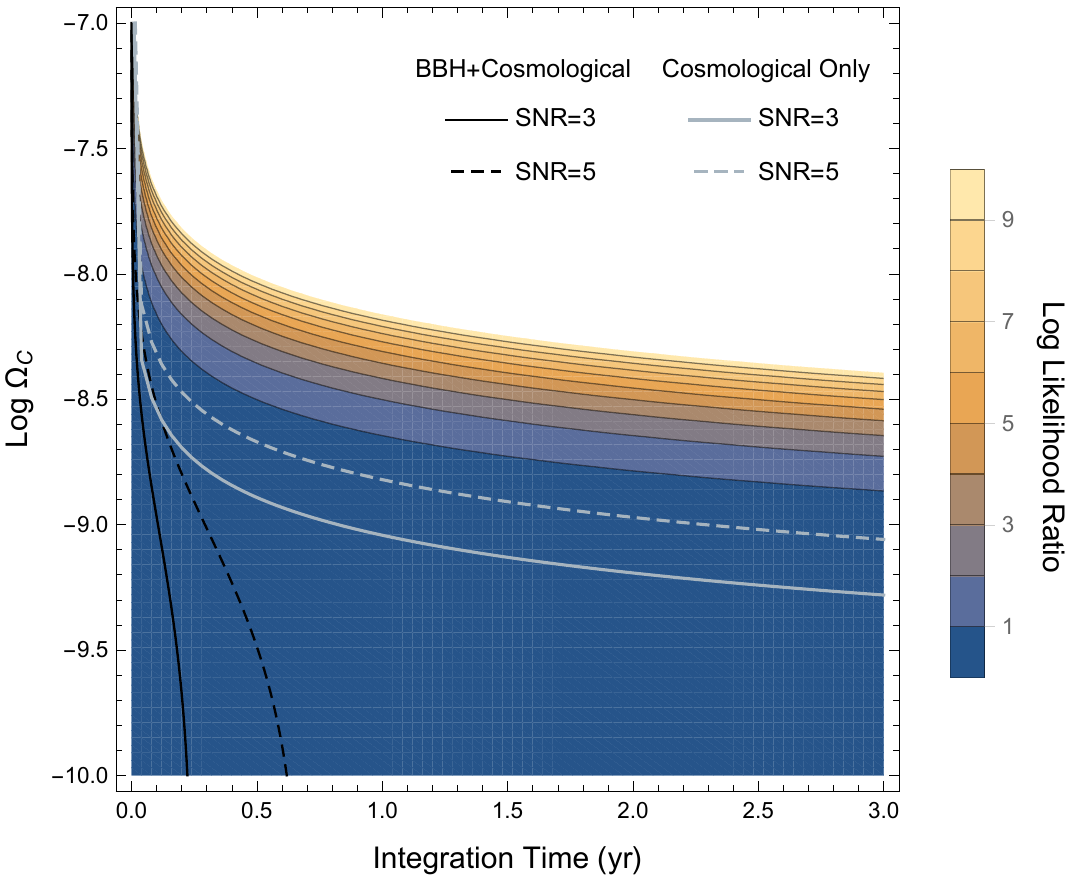}
  \caption{Contours of the maximum likelihood ratio between the ``BBH+'' and ``BBH-only'' models, as a function of the amplitude of the cosmological background $\Omega_c$ and the total integration time.
  Results are shown for both the H1-L1 detector combination (\textit{left}) and for two colocated detectors (\textit{right}).
  Black curves indicate observation times required to detect the combined (astrophysical plus cosmological) background with a given optimal SNR.
  Gray curves indicate the amplitude of a flat cosmological background that alone would be detectable to a given SNR. Solid curves represent $\text{SNR}=3$ and dashed curves represent $\text{SNR}=5$.  }
  \label{fig:cosmologicalLikelihoods}
\end{figure*}

Contours of the maximum log-likelihood ratio are shown in Fig.~\ref{fig:cosmologicalLikelihoods} as a function of the cosmological background amplitude $\Omega_c$ and the total integration time, for the H1-L1 detector network (left-hand panel) and for two colocated detectors (right-hand panel).
In each panel, the black solid (dashed) curves indicate the observation times necessary to detect the combined astrophysical and cosmological background with optimal $\text{SNR}=3$ $(5)$; note that these curves become vertical as $\Omega_c$ approaches zero, corresponding to the fixed detection time of the BBH background alone.
The gray solid (dashed) curves indicate the cosmological background amplitudes that would otherwise be detectable with optimal $\text{SNR}=3$ $(5)$, if there existed no BBH background.

The fact that the gray curves lie deep within the $\ln\mathcal{R} \simeq 0$ region implies that the presence of a BBH background serves to obscure any cosmological background that would otherwise be detectable.
If no BBH background were present, for instance, Advanced LIGO could detect a cosmological background of amplitude $\Omega_c \approx 10^{-9.0}$ with $\text{SNR}=3$ after three years of observation.
When a BBH background is present, however, a much larger amplitude of $\Omega_c \approx 10^{-8.2}$ (corresponding to $\ln\mathcal{R}=3$) is required to resolve an additional flat background component.
After three years of observation at design sensitivity, Advanced LIGO will therefore be able to constrain the amplitudes of additional background components to $\Omega_c \lesssim 10^{-8.2}$.
A network of colocated detectors performs somewhat better, constraining additional background components to $\Omega_c \lesssim 10^{-8.4}$ after one year of observation and to $\lesssim 10^{-8.6}$ after three years of observation.

In the above, we treat the power-law amplitudes $\Omega_0$ and $\Omega_1$ of Eqs.~\eqref{bbhOnly} and \eqref{bbhPlus} as entirely free parameters.
In reality, we will likely be able to place a prior on these parameters, using CBC estimates of the average chirp mass and local coalescence rate as well as estimates of the theoretical uncertainty in background modeling~\cite{gw150914_stoch}.
However, even if we assume the amplitudes $\Omega_0$ and $\Omega_1$ are known to within a factor of 2 (an optimistic assumption given the uncertainty in the merger rate evolution with redshift~\cite{gw150914_stoch}), we find little change in the results presented in Fig.~\ref{fig:cosmologicalLikelihoods}.
With optimistic priors, the ability of H1-L1 and H1-H2 to resolve a cosmological background is improved only for observation times $T\lesssim1$ yr.
After $T\approx1$ yr of integration, the experimental uncertainty on the stochastic background amplitude falls below the width of the prior distribution, and so the priors are no longer informative.

\section{CONCLUSION AND FUTURE OUTLOOK}

In this paper, we seek to address three questions concerning an astrophysical binary black hole background. First, how does the information contained in a stochastic background compare with what can be learned from nearby, individually resolvable binary mergers?
Direct searches for binary black hole coalescences are, on average, sensitive to redshifts less than $z_{50\%}\approx0.5$ for events like GW150914.
The stochastic background, on the other hand, is dominated by binary mergers in the far more distant Universe, with 90\% of the stochastic SNR due to sources between redshifts $z\approx0.1$ and $3.5$.
The stochastic background therefore encodes astrophysical information (e.g., the mass distribution and rate of BBH mergers as a function of redshift) about a population of black hole binaries that is distinct from the local population visible to CBC searches.

Second, what astrophysics can we hope to extract from future observations of the binary black hole background?
In principle, the functional form of the background's energy density spectrum depends upon the precise characteristics of the underlying binary black hole population (mean chirp mass, local coalescence rate, star formation history, etc).
We demonstrate, however, that for realistic chirp masses and coalescence rates, the form of the stochastic background is indistinguishable from a simple $f^{2/3}$ power law with Advanced LIGO.
In the near future, parameter estimation and model selection on the stochastic background are therefore limited to measuring only the overall amplitude of the background.

Finally, how is our ability to measure other stochastic backgrounds affected by the presence of an astrophysical BBH background?
We find that an astrophysical BBH background obscures the presence of any underlying cosmological background that might otherwise be detected with Advanced LIGO.
For such a cosmological background to be resolvable, it must be strong enough to overcome our uncertainty in the amplitude of the BBH background.
In this sense, the BBH background now acts as a foreground, limiting Advanced LIGO's sensitivity to additional, weaker background components.

It should be noted that the \texttt{Fiducial} and \texttt{LowMetallicity} background models we consider here make specific assumptions about the metallicities of black hole progenitors and the masses and formation times of black hole binaries, properties that are currently only poorly understood.
Different models of the BBH background will yield different numerical results for the above three questions.
Qualitatively, however, the above conclusions are robust.

Future developments, however, may brighten these prospects.
It may be possible to achieve better model selection than we show here through the development of a non-Gaussian stochastic pipeline, optimized for signals like the expected BBH background.
Future stochastic measurements will also be strongly aided by any developments in instrumentation or data analysis that improve detector sensitivities at high frequencies, as it is only at high frequencies that the BBH background deviates substantially from a power law.
To this end, one strategy is the use of colocated detectors, such as the H1-H2 configuration of Initial LIGO, to avoid the penalty associated with the overlap reduction function at high frequencies (at the potential cost of introducing correlated environmental noise)~\cite{H1H2_2014}.
As see in Sec. \ref{BBHbackground}, it may in fact be possible for an Advanced LIGO H1-H2 configuration to differentiate between astrophysical background models on a more practical time scale of $\sim5-10$ yr (as opposed to hundreds or thousands of years with the H1-L1 configuration).

In the more distant future, third-generation detectors like the Einstein Telescope (ET)~\cite{cqg.27.194002.10} will be able to probe black hole binaries at cosmological distances.
ET is projected to resolve individual events like GW150914 to redshifts of $z\sim15$~\cite{ET2013}, allowing for precision observation of the binary black hole population over the entire history of star formation.
The ability of ET to resolve such events raises the exciting possibility of the individual identification and subtraction of each BBH coalescence from the data, opening the way for the detection of weaker, underlying stochastic backgrounds of astrophysical or even cosmological origin.

\section*{ACKNOWLEDGEMENTS}
We thank Christopher Berry, Nelson Christensen, Eric Howell, Vuk Mandic, Duncan Meacher, Tania Regimbau, and Alan Weinstein.
T. C. is a member of the LIGO Laboratory, supported by funding from the U.S. National Science Foundation. LIGO was constructed by the California Institute of Technology and Massachusetts Institute of Technology with funding from the National Science Foundation and operates under cooperative agreement Grant No. PHY-0757058.
ET is supported by ARC FT150100281. IM was partially supported by STFC and by the Leverhulme Trust.

This paper has been assigned the LIGO document number LIGO-P1600059.

\appendix

\section{BINARY BLACK HOLE ENERGY SPECTRUM}
\label{bbhappendix}

We adopt the BBH model described in Ref.~\cite{Ajith_EA_2008}, which presents a phenomenological description of the inspiral, merger, and ringdown of a black hole coalescence.
The corresponding energy spectrum for a single binary is~\cite{2011ApJ...739...86Z}
	\begin{equation}
	\frac{dE\bbh}{df} = \frac{\left(G\pi\right)^{2/3} \mathcal{M}_c^{5/3}}{3} H(f),
	\end{equation}
where
	\begin{equation}
	H(f) = \begin{cases}
		f^{-1/3} & (f < f_\text{merge}) \\
		\frac{f^{2/3}}{f_\text{merge}} & (f_\text{merge} \leq f < f_\text{ring}) \\
		\frac{1}{f_\text{merge} f_\text{ring}^{4/3}} \left( \frac{f}{ 1 + \left( \frac{f - f_\text{ring}}{\sigma/2}\right)^2 } \right)^2 & (f_\text{ring} \leq f < f_\text{cutoff}) \\
		0 & (f \geq f_\text{cutoff})
		\end{cases}.
	\end{equation}
Definitions for $f_\text{merge}$, $f_\text{ring}$, $f_\text{cutoff}$, and $\sigma$ are given in Sec. IV of Ref.~\cite{Ajith_EA_2008}.

\section{STAR FORMATION AND MEAN METALLICITY EVOLUTION}
\label{sfrappendix}

Following the \texttt{Fiducial} model of Ref.~\cite{gw150914_stoch}, we adopt the star formation rate~\cite{2015MNRAS.447.2575V,Kistler2013}
	\begin{equation}
	R_*(z) = \nu \frac{ a \exp\left[b(z-z_m)\right] }{ a - b + b \exp\left[a(z-z_m)\right] }\,\frac{M_\odot}{\text{Mpc}^{3}\,\text{yr}},
	\end{equation}
with $\nu = 0.145$, $z_m = 1.86$, $a = 2.80$, and $b = 2.62$.

In the \texttt{Fiducial} model, the rate of binary black hole formation is proportional to the rate of star formation with metallicity below $Z_\odot/2$.
The mean stellar metallicity is given as a function of redshift by~\cite{gw150914_stoch,Madau_Dickinson_2014}
	\begin{equation}
	\overline{\log Z(z)} = 0.5 + \log \left( \frac{y(1-R)}{\rho_b} \int_z^{20} \frac{ R^\textsc{md}_*(z') dz'}{H(z') (1+z')} \right),
	\end{equation}
with stellar metal yield $y = 0.019$, return fraction $R = 0.27$, baryon density $\rho_b = 2.77\times10^{11} \Omega_b \,h_0^2\,M_\odot\,\text{Mpc}^{-3}$, and $\Omega_b = 0.045$.
The star formation rate used in calibrating $y$ and $R$ is~\cite{Madau_Dickinson_2014}
	\begin{equation}
	R^\textsc{md}_*(z) = 0.015 \frac{ (1+z)^{2.7} }{ 1 + \left(\frac{1+z}{2.9}\right)^{5.6} }\,\frac{M_\odot}{\text{Mpc}^{3}\,\text{yr}}.
	\end{equation}
Assuming that stellar metallicity is log-normally distributed with a standard deviation of $0.5$, the fraction of stars with $Z < Z_\odot/2$ is
	\begin{equation}
	F(z) = \frac{ 
		\int_{-\infty}^{\log Z_\odot/2} \exp \left\{ - 2 \left[ \log Z - \overline{\log Z(z)} \right]^2 \right\} d\log Z
		}{
		\int_{-\infty}^\infty \exp \left\{ - 2 \left[ \log Z - \overline{\log Z(z)} \right]^2 \right\} d\log Z
		}.
	\end{equation}
The rate of binary black hole formation is assumed to be proportional to $R_*(z) F(z)$.

\section{THRESHOLD REDSHIFTS}
\label{rangeappendix}

The optimal signal-to-noise ratio of a single-detector matched filter search is
	\begin{equation}
	\rho^2 = 4 \int_0^\infty \frac{ | \tilde h(f) |^2 }{ P(f)} df,
	\end{equation}
where $\tilde h(f)$ is the measured strain signal.
Using the phenomenological BBH model of Ref.~\cite{Ajith_EA_2008}, the signal-to-noise ratio of an optimally positioned and oriented binary is
	\begin{equation}
	\label{cbcSNR}
	\rho^2 = \frac{5}{6} \frac{c^2}{\pi^{4/3} D_L^2} \left(\frac{G \mathcal{M}_c (1+z) } {c^3}\right)^{5/3} f_\text{merge}^{-7/3} 
		\int_0^\infty \frac{s(f)^2}{P(f)} df,
	\end{equation}
where $D_L = D(1+z)$ is the luminosity distance, $D$ the proper distance to the source, and
	\begin{equation}
	s(f) = \begin{cases}
		\left( \frac{f}{f_\text{merge}}\right)^{-7/6} & (f < f_\text{merge}) \\
		\left( \frac{f}{f_\text{merge}}\right)^{-2/3} & (f_\text{merge} \leq f < f_\text{ring}) \\
		\left( \frac{f_\text{ring}}{f_\text{merge}} \right)^{-2/3} \frac{ \sigma^2/4}{ \left(f-f_\text{ring}\right)^2 + \sigma^2/4} 
			& ( f_\text{ring} \leq f < f_\text{cutoff} \\
		0 & (f \geq f_\text{cutoff})
		\end{cases}.
	\end{equation}
The values of $f_\text{merge}$, $f_\text{ring}$, $f_\text{cutoff}$, and $\sigma$ are given in Ref.~\cite{Ajith_EA_2008}.
The source distance $D$ is given in terms of redshift by
	\begin{equation}
	D(z) = c \int_0^z \frac{dz'}{H(z')}.
	\label{distEq}
	\end{equation}

In general, the squared signal-to-noise ratio of an arbitrarily positioned and oriented source is reduced from the optimal value Eq.~\eqref{cbcSNR} by an antenna factor $\mathcal{F}$, which depends on the source's sky position, polarization angle, and inclination.
Given an ensemble of randomly positioned and oriented sources, the median value of $\mathcal{F}$ is $\langle\mathcal{F}\rangle_\text{med} = 0.107$.
The threshold redshifts $z_{50\%}$ quoted in Sec. \ref{BBHbackground} are obtained by numerically solving Eqs.~\eqref{cbcSNR} and~\eqref{distEq} for the redshift at which the squared signal-to-noise ratio of an optimally positioned and oriented binary is $\rho^2 = 64/\langle\mathcal{F}\rangle_\text{med}$.
Beyond redshift $z_{50\%}$, less than $50\%$ of binaries are directly resolvable.

\end{document}